\documentclass[amssymb,prd,aps,epsfig,showpacs,nofootinbib]{revtex4}

\newcommand{\be}{\begin{equation}}
\newcommand{\ee}{\end{equation}}
\newcommand{\ba}{\begin{eqnarray}}
\newcommand{\ea}{\end{eqnarray}}

\begin{document}
\title{{\bf \ Supersymmetric Extension of the Lorentz and CPT-violating
Maxwell-Chern-Simons Model}}
\author{H. Belich$^{a,b}$, J. L. Boldo$^{b,d}$, L. P. Colatto$^{a,b}$, J.A.
Helay\"{e}l-Neto$^{a,b}$ and A.L.M.A. Nogueira$^{a,b}$\thanks{{\tt e-mails:} 
{\tt belich@cbpf.br, jboldo@cce.ufes.br, colatto@cbpf.br, helayel@cbpf.br,
nogue@cbpf.br}}}
\affiliation{$^{a}${\it Centro Brasileiro de Pesquisas F\'{i}sicas (CBPF)},\\
Coordena\c{c}\~{a}o de Teoria de Campos e Part\'{i}culas (CCP), \\
Rua Dr. Xavier Sigaud, 150 - Rio de Janeiro - RJ 22290-180 - Brazil.\\
$^{b}${\it Grupo de F\'{i}sica Te\'{o}rica Jos\'{e} Leite Lopes, }\\
Petr\'{o}polis - RJ - Brazil. \\
$^{d}$ {\it Universidade Federal do Esp\'{i}rito Santo (UFES)},\\
Departamento de F\'{i}sica e Qu\'{i}mica, Av. Fernando Ferrarim, S/N\\
Goiabeiras, Vit\'{o}ria - ES, 29060-900 - Brasil}

\begin{abstract}
\noindent
Focusing on gauge degrees of freedom specified by a 1+3 dimensions model
hosting a Maxwell term plus a Lorentz and CPT non-invariant
Chern-Simons-like contribution, we obtain a minimal extension of such a
system to a supersymmetric environment. We comment on resulting peculiar
self-couplings for the gauge sector, as well as on background contribution
for gaugino masses. Furthermore, a non-polynomial generalization is
presented.
\end{abstract}

\maketitle

\section{\ Introduction\-}

Lorentz and CPT invariances are cornerstones in modern Quantum Field Theory,
both symmetries being respected by the Standard Model for Particle Physics.
Nevertheless, nowadays one faces the possibility that this scenario is only
an effective theoretical description of a low-energy regime, an assumption
that leads to the idea that these fundamental symmetries could be violated
when one deals with energies close to the Planck scale \cite{Jackiw}. Taking
this viewpoint, several approaches to analyze the violation of Lorentz
symmetry have been proposed in the literature. Eventually a common feature
arises: the violation is implemented by keeping either a four-vector (in a
CPT-odd term \cite{Jackiw}) or a traceless symmetric tensor (CPT-preserving
term \cite{Kostelec1}) unchanged by particle inertial frame transformations 
\cite{Colladay} which is generally called spontaneous violation.
Furthermore, the issue of preserving supersymmetry ({\it Susy}) while
violating Lorentz symmetry is addressed to \cite{berger}. This breaking of
Lorentz symmetry is also phenomenologically motivated as a candidate to
explain the patterns observed in the detection of ultra-high energy cosmic
rays, concerning the events with energy above the GZK ( $E_{GZK}\simeq
4\times 10^{19}~eV.T$) cutoff \cite{raios}. Moreover, measurements of radio
emission from distant galaxies and quasars verify that the polarization
vectors of these radiations are not randomly oriented as naturally expected.
This peculiar phenomenon suggests that the space-time intervening between
the source and observer may be exhibiting some sort of optical activity, the
origin of which is not known.

In a Theoretical Field proposal where this breaking of Lorentz invariance is
taken into account, an analysis of the unitarity, causality, and vortex-like
solutions had been carried out in Ref. \cite{Baeta}. Another focus of
interest points to planar gauge systems, which play a relevant role in
Condensed Matter descriptions, as they happen to be related to issues like
high-Tc superconductivity and fractionary quantum Hall effect. Possible
contributions from Lorentz-violating terms to the appearance of anisotropy
in planar systems had been investigated in Refs. \cite{Manojr} and \cite
{beto}.

A first proposal of Supersymmetry-Preserving Lorentz Violation was carried
out in the work of Ref.\cite{berger}. The aim of that work was to
investigate whether one could maintain desired properties of supersymmetric
systems, namely, cancellation of divergences and the patterns of spontaneous
breaking schemes, while violating Lorentz symmetry. A Lorentz breaking
tensor with constants entries has been adopted following an original
suggestion given by Colladay \cite{Colladay}. Working upon a modified
Wess-Zumino model, the authors of Ref. \cite{berger} had demonstrated that
convenient changes of the Susy-algebra of fermionic charges and of
Susy-covariant derivatives expressions were enough to define a Susy-like
invariance for the Lorentz violating starting theory. As a matter of fact
the modification of the algebra was achieved by adding a particular
tensor-dependent central term, of the $k_{\mu \upsilon }\partial ^{\nu }$
type, where $k_{\mu \upsilon }$ exhibits real symmetric traceless tensor
properties.

As a net result, it was shown that a model for a modified-Susy invariant but
Lorentz non-invariant {\it matter} system can be built. Moved by a different
perspective, we now present an analysis on Lorentz and Susy breakings
concerning degrees of freedom in the {\it gauge} field sector. We start off
by establishing the supersymmetric minimal extension for the
Chern-Simons-like term \cite{Jackiw}, 
\begin{equation}
\Sigma _{CS}=-\frac{1}{4}\int dx^{4}\epsilon ^{\mu \nu \alpha \beta }c_{\mu
}A_{\nu }F_{\alpha \beta },  \label{jac}
\end{equation}
preserving the usual $(1+3)$-dimensional Susy algebra. The breaking of Susy
will follow the very same route to Lorentz breaking: the statement that $%
c_{\mu }$ is a constant (in the active sense) vector triggers both Lorentz
and, as we shall comment on, Susy breakings. Handling proper superfield
extensions for the background shall prevent the model from displaying higher
spin excitations, and interesting self-couplings for the gauge sector as
well as background contribution for the gaugino masses come up naturally as
a consequence of the (initially) supersymmetric structure.

In the next section, we present the Susy minimal extension for \ref{jac}. In
Section 2, a first generalization, with non-polynomial couplings, shows up.
Finally, we comment on conclusions and perspectives in Section 4.

\section{The Supersymmetric Extension of the Maxwell-Chern-Simons Model.}

Adopting covariant superspace-superfield formulation, we propose the
following minimal extension for \ref{jac}: 
\begin{equation}
A=\int d^{4}xd^{2}\theta d^{2}\bar{\theta}\left\{ W^{a}(D_{a}V)S+\overline{W}%
_{\dot{a}}(\overline{D}^{\dot{a}}V)\overline{S}\right\} ,  \label{superjac}
\end{equation}
where the superfields $W_{a}$, $V$, $S$ and the Susy-covariant derivatives $%
\ D_{a}$, $\overline{D}_{\dot{a}}$ hold the definitions: 
\begin{eqnarray*}
D_{a} &=&\frac{\partial }{\partial \theta ^{a}}+i{\sigma ^{\mu }}_{a\dot{a}}%
\bar{\theta}^{\dot{a}}\partial _{\mu } \\
\overline{D}_{\dot{a}} &=&-\frac{\partial }{\partial \bar{\theta}^{\dot{a}}}%
-i\theta ^{a}{\sigma ^{\mu }}_{a\dot{a}}\partial _{\mu };
\end{eqnarray*}
from ${\overline{D}}_{\dot{b}}W_{a}\left( x,\theta ,\bar{\theta}\right) =0$
and $D^{a}W_{a}\left( x,\theta ,\bar{\theta}\right) =$ $\overline{D}_{\dot{a}%
}\overline{W}^{\dot{a}}\left( x,\theta ,\bar{\theta}\right) $, it follows
that
\[
W_{a}(x,\theta ,\bar{\theta})=-\frac{1}{4}\overline{D}^{2}D_{a}V:
\]
Its $\theta $-expansion reads as below:
\begin{eqnarray*}
W_{a}(x,\theta ,\bar{\theta}) &=&\lambda _{a}\left( x\right) +i\theta ^{b}{%
\sigma ^{\mu }}_{b\dot{a}}\bar{\theta}^{\dot{a}}\partial _{\mu }\lambda
_{a}\left( x\right) -\frac{1}{4}{\bar{\theta}}^{2}\theta ^{2}\square \lambda
_{a}\left( x\right)  \\
&&+2\theta _{a}D\left( x\right) -i\theta ^{2}\bar{\theta}^{\dot{a}}{\sigma
^{\mu }}_{a\dot{a}}\partial _{\mu }D\left( x\right)  \\
&&+{{{\sigma }^{\mu \nu }}_{a}}^{b}\theta _{b}F_{\mu \nu }\left( x\right) -%
\frac{i}{2}{{{\sigma }^{\mu \nu }}_{a}}^{b}{\sigma ^{\alpha }}_{b\dot{a}%
}\theta ^{2}\overline{\theta }^{\dot{a}}\partial _{\alpha }F_{\mu \nu
}\left( x\right)  \\
&&-i\sigma _{a\dot{a}}^{\mu }\partial _{\mu }\text{ }\overline{\lambda }^{%
\dot{a}}\left( x\right) \theta ^{2}
\end{eqnarray*}
and $V=V^{\dagger }$. The Wess-Zumino gauge choice is taken as usually done:
\[
\text{ }V_{WZ}=\theta \sigma ^{\mu }\bar{\theta}A_{\mu }(x)+\theta ^{2}\bar{%
\theta}\overline{\lambda }\left( x\right) +\bar{\theta}^{2}\theta \lambda
(x)+\theta ^{2}\bar{\theta}^{2}D,
\]
so the\ action (\ref{superjac}) is gauge-invariant. The background
superfield is so chosen to be a chiral one. Such a constraint restricts the
maximum spin component of the background to be an $s$ $=$ $\frac{1}{2}$
component-field, showing up as a Susy-partner for a spinless dimensionless
scalar field. Also, one should notice that $S$ happens to be dimensionless.
The superfield expansion for$\ S$ then reads:
\begin{eqnarray*}
\overline{D}_{\dot{a}}S\left( x\right)  &=&0\text{ \ and }S\left( x\right) 
\text{ }=s\left( x\right) +i\theta \sigma ^{\mu }\overline{\theta }\partial
_{\mu }s\left( x\right) -\frac{1}{4}{\bar{\theta}}^{2}\theta ^{2}\square
s\left( x\right)  \\
&&+\sqrt{2}\theta \psi \left( x\right) +\frac{i}{\sqrt{2}}\theta ^{2}%
\overline{\theta }\overline{\sigma }_{\mu }\partial _{\mu }\psi \left(
x\right) +\theta ^{2}F\left( x\right) .
\end{eqnarray*}
The component-wise counterpart for the action (\ref{superjac}) is as
follows: \bigskip 
\begin{eqnarray}
A_{comp.} &=&\int d^{4}x\text{ }\left\{ -\frac{1}{2}(s+s^{\ast })F_{\mu \nu
}F^{\mu \nu }+\frac{i}{2}\partial _{\mu }(s-s^{\ast })\varepsilon ^{\mu
\alpha \beta \nu }F_{\alpha \beta }A_{\nu }+4D^{2}(s+s^{\ast })\right. 
\nonumber \\
&-&2is\,\lambda \,\sigma ^{\mu }\partial _{\mu }\overline{\lambda }%
-2is^{\ast }\,\overline{\lambda }\,\overline{\sigma }^{\mu }\partial _{\mu
}\lambda -\sqrt{2}\lambda (\sigma ^{\mu \nu })F_{\mu \nu }\psi +\sqrt{2}%
\overline{\lambda }(\overline{\sigma }^{\mu \nu })F_{\mu \nu }\overline{\psi 
}+  \nonumber \\
&+&\left. \lambda \,\lambda F+\overline{\lambda }\,\overline{\lambda }%
F^{\ast }-2\sqrt{2}\lambda \,\psi D-2\sqrt{2}\overline{\lambda }\,\overline{%
\psi }D\right\}  \label{comps}
\end{eqnarray}
\bigskip As one can easily recognize, the first line displays the 4D
Chern-Simons-like term (\ref{jac}), where the vector $c_{\mu }$ is expressed
as the gradient of a real background scalar: $c_{\mu }$ $=$ $\partial _{\mu
}\sigma $, for $s$ $=$ $\xi +i\sigma $. Such a reduction of the vector into
a gradient of a scalar field stems directly from the simultaneous
requirements of both gauge\footnote{%
The gauge invariance of action \ref{superjac} will become clearly manifest
in the next section, where we rephrase the supersymmetrization of the 4D
Chern-Simons-like term in a formulation restricted to the chiral
(anti-chiral for the h.c. counterpart) sector of superspace.} and
supersymmetry invariances.

Another interesting feature of this model concerns the presence of
self-couplings for the gauge sector: the fermionic background field, $\psi $%
, triggers the coupling of the gauge boson (through the field-strength) to
the gaugino. Moreover, using the field equation for the gauge auxiliary
field $D$ one arrives at a quartic fermionic fields coupling - $\lambda
\lambda \psi \psi $ -, and the background nature of $\psi $ indicates a
background contribution for the gaugino mass\footnote{%
We shall analyze the propagator structure for the gauge component-fields in
a forthcoming communication. We anticipate that a constant $\psi $
component-field configuration is compatible with the supersymmetry algebra.}.

Concerning the breaking of Lorentz symmetry, realized by assuming $c_{\mu }$ 
$=$ $\partial _{\mu }\sigma $ to be constant under the action of particle
inertial frame transformations, one should observe that such an assumption
implies that the scalar component-field $\sigma $ must be linear in the
coordinates, $\sigma =c_{\mu }x^{\mu }$. As a matter of fact, a linear
dependence on $x^{\mu }$ cannot be implemented by means of a Susy-covariant
constraint (i.e., Susy-covariant derivatives acting on $S$), and, in that
sense, the choice of a rigid $\partial _{\mu }\sigma $ breaks Susy in exact
analogy to the Lorentz breaking scheme adopted. To better establish such a
correspondence, one can consider the choice for constant $\partial _{\mu
}\sigma $ to be accompanied by a constant $\psi $ requirement (and a
constant auxiliary field, $F$, as well\footnote{%
In fact, a constant auxiliary field $F$ is singled out as a susy-invariant
parameter, as far as one deals with a constant $\psi $.}). In this context,
a (passive) Susy-transformation keeps the status of all component-fields
unchanged.

In the next section, we provide the model with a non-polynomial
generalization, which brings about the possibility of understanding the 4D
C.S.-like term as a first order correction in a complete exponential
scenario.

\section{Non-polynomial generalization}

\bigskip

Let us note that the integration defined through the Grassmanian measure $%
d^{2}\bar{\theta}$ (or $d^{2}\theta $ ) can be represented by the action of
a squared Susy-covariant derivative (up to a normalization factor), $%
\overline{D}^{2}$ (or $D^{2}$), on the super-Lagrangian $W^{a}(D_{a}V)S$ $+$ 
$h.c.$, if one neglects boundary terms, and that the only sector of the
superfield product $W(DV)S$\ (or $\overline{W}(\overline{D}V)\overline{S}$)
that admits a non-null action of $\overline{D}^{2}$ (or $D^{2}$) is the
factor $DV$ (or $\overline{D}V$ ). Such a manipulation leads to the
Lagrangian $d^{4}x($ $d^{2}\theta $ $W^{a}(\overline{D}^{2}D_{a}V)S$ $+$ $%
d^{2}\bar{\theta}$ $\overline{W}_{\dot{a}}(D^{2}\overline{D}^{\dot{a}}V)%
\overline{S}$\bigskip $)$, and one can rephrase (\ref{superjac}) through
such a parametrization: 
\[
A=h\int d^{4}x\left\{ d^{2}\theta \lbrack W^{a}W_{a}S]+d^{2}\bar{\theta}%
\left[ \overline{W}_{\dot{a}}\overline{W}^{\dot{a}}\overline{S}\right]
\right\} ,
\]
where a suitable dimensionless (perturbation) parameter $h$ is inserted. We
remark that such an inclusion does not spoil any power-counting
renormalization property of the model. Moreover, as we aim at a Susy version
for a model hosting both regular Maxwell kinetic term and the 4D C.S.-like
term \cite{Baeta}, we end up with the following combination: 
\[
A_{Max.+C.S.}=\frac{1}{4}\int d^{4}x\left\{ d^{2}\theta \lbrack
W^{a}W_{a}]+d^{2}\bar{\theta}\left[ \overline{W}_{\dot{a}}\overline{W}^{\dot{%
a}}\right] \right\} +\frac{h}{4}\int d^{4}x\left\{ d^{2}\theta \lbrack
W^{a}W_{a}S]+d^{2}\bar{\theta}\left[ \overline{W}_{\dot{a}}\overline{W}^{%
\dot{a}}\overline{S}\right] \right\} .
\]
Such an expression induces a straightforward non-polynomial generalization: 
\begin{equation}
A_{non-pol.}=\frac{1}{4}\int d^{4}x\left\{ d^{2}\theta \left[ W^{a}W_{a}\exp
(hS)\right] +d^{2}\bar{\theta}\left[ \overline{W}_{\dot{a}}\overline{W}^{%
\dot{a}}\exp (h\overline{S})\right] \right\} ,  \label{expon}
\end{equation}
leaving room for a perturbative approach parametrized by orders of $h$. In
fact, the action (\ref{expon}) includes a zero order supersymmetric Maxwell
theory, a first-order Susy-extended 4D C.S.-like term (reproducing the
action of the eq. (\ref{comps})), and higher orders contributions. In
component-field parametrization, action (\ref{expon}) reads:
\begin{eqnarray*}
A_{non-pol.} &=&\frac{1}{4}\int d^{4}x\left\{ \exp (hs)\left[ -\frac{1}{2}%
F_{\mu \nu }F^{\mu \nu }-\frac{i}{2}\widetilde{F}_{\mu \nu }F^{\mu \nu
}-2i\lambda ^{a}{\sigma ^{\mu }}_{a\dot{a}}\partial _{\mu }\text{ }\overline{%
\lambda }^{\dot{a}}+4D^{2}+\right. \right. \\
&&+\left. \left. h\left( -2\sqrt{2}\lambda ^{a}\psi _{a}D+\lambda
^{a}\lambda _{a}F-\sqrt{2}\lambda ^{a}\left. (\sigma ^{\mu \nu })_{a}\right.
^{b}F_{\mu \nu }\psi _{b}\right) -\frac{h^{2}}{2}\lambda ^{a}\lambda
_{a}\psi ^{b}\psi _{b}\right] +h.c..\right\}
\end{eqnarray*}
The exponential version brings about the 4D C.S.-like term in the form $-%
\frac{i}{8}\exp (hs)\widetilde{F}_{\mu \nu }F^{\mu \nu }+h.c.$, demanding an
integration by parts to reproduce the expression $i\partial _{\mu
}(s-s^{\ast })\varepsilon ^{\mu \alpha \beta \nu }F_{\alpha \beta }A_{\nu }$%
. One should also realize that a quartic fermion-fields coupling is already
present at order $h^{2}$ , even if the field equation for the auxiliary
field $D$ is not used to eliminate it. It is also interesting to observe how
the background components $s$, $\psi $ and $F$ influence on the gaugino
physical mass.

\section{Concluding Comments}

Working on the {\it gauge}-field sector of a system with a Lorentz breaking
4D-Chern-Simons-like term, we have been able to derive its minimal
supersymmetric extension and a peculiar nom-polynomial generalizations has
been proposed that is compatible with $N=1$-Susy. Focusing on the minimal
Susy-extension, one should already realize the presence of new couplings
induced by the background (passive-)superfield components. The assumption
that the Lorentz breaking is implemented by means of a constant vector,
regarded as a background input, finds its as a Susy-counterpart in a set of
requirements on the space-time dependence of each component-field of the
background superfield, $S$. A scalar field, $s$, linearly dependent on $%
x^{\mu }$, as well as a constant spinor field, $\psi $, arise as a
consequence of gauge invariance, and these results impose that, eventually,
coupling terms are to be regarded as mass terms. A complete analysis of the
propagator structure for the gauge supermultiplet, both in superspace and in
component-fields, is mandatory, including an interesting study of the
gaugino (background-)induced mass. We shall very soon report our efforts in
this matter elsewhere.

\section{Acknowledgments}
The authors are grateful to A.P. Ba\^eta Scarpelli for several discussions.
One of the authors, J.L.B., would like to thank CCP/CBPF for the hospitality.

\end{document}